\documentclass[showpacs,preprintnumbers,amsmath,amssymb,twocolumn,superscriptaddress,prb]{revtex4}
\usepackage{times}
\usepackage{graphicx}
\usepackage{amsfonts}
\usepackage{amsmath, amsthm, amssymb}
\usepackage{dsfont}

\newcommand{\vk}{\mathbf{k}} 
\newcommand{\e}[1]{\mathrm{e}^{#1}}
\newcommand{\cop}{\hat{c}} 

\newcommand{\ie}{\textit{i.e. }}
\newcommand{\eg}{\textit{e.g. }}
\newcommand{\etal}{\emph{et al.}}
\def\i{\mathrm{i}}                            
\newcommand{\kk}{\mathbf{k}}
\newcommand{\dd}{\mathbf{d}}
\newcommand{\be}{\begin{equation}}
\newcommand{\ee}{\end{equation}}

\begin{document}
\title[Coexistence of itinerant ferromagnetism and a non-unitary superconducting state with line nodes: possible 
application to UGe$_2$]{Coexistence of itinerant ferromagnetism and a non-unitary superconducting state with 
line nodes: possible application to UGe$_2$}
\author{Jacob Linder}
\affiliation{Department of Physics, Norwegian University of
Science and Technology, N-7491 Trondheim, Norway.}
\author{Iver B. Sperstad}
\affiliation{Department of Physics, Norwegian University of
Science and Technology, N-7491 Trondheim, Norway.}
\author{Andriy H. Nevidomskyy}
\affiliation{Department of Physics and Astronomy,
Rutgers University, Piscataway, N. J., 08854-8019}
\author{Mario Cuoco}
\affiliation{Laboratorio Regionale SuperMat, CNR-INFM Salerno, Baronissi (Sa), Italy}
\affiliation{Dipartimento di Fisica "E.R. Caianiello", Universit´a di Salerno, Baronissi(Sa),Italy}
\author{Asle Sudb{\o}}
\affiliation{Department of Physics, Norwegian University of
Science and Technology, N-7491 Trondheim, Norway.}
\date{Received \today}
\begin{abstract}
We construct a mean-field theory for itinerant ferromagnetism
coexisting with a non-unitary superconducting state, where only the
majority-spin band is gapped and contains line nodes, while the
minority-spin band is gapless at the Fermi level. Our study is
motivated by recent experimental results indicating that this may be
the physical situation realized in the heavy-fermion compound
UGe$_2$. We investigate the stability of the mean-field solution of
the magnetic and superconducting order parameters. Also, we provide
theoretical predictions for experimentally measurable properties of
such a non-unitary superconductor: the specific heat capacity,
the Knight shift, and the tunneling conductance
spectra. Our study should be useful for direct comparison with experimental 
results and also for further predictions of the physics that may be expected 
in ferromagnetic superconductors.
  \end{abstract}

\maketitle

\section{Introduction}
The interplay between ferromagnetic
(FM) and superconducting (SC) long range order microscopically coexisting in the same 
material has attracted much interest during the last decade due to the discovery of 
superconductivity in ferromagnetic metals, UGe$_2$, URhGe, UCoGe, and possibly ZrZn$_2$ 
(see, however, Ref. \onlinecite{yelland}).\cite{saxena, aoki, huy, ohta, pfleiderer}. 
One possible route of investigation of such systems was adopted in
early works~\cite{abrikosov,clogston,chandrasekhar}, which assumed a
conventional $s$-wave superconducting condensate residing in a
ferromagnetic background caused by localized spins or aligned magnetic
impurities. It was shown that below a critical value of the magnetic
coupling, comparable to the superconducting gap $\Delta$ itself, 
superconductivity and magnetism were able to coexist. 
 It was also suggested that a finite momentum pairing state, known as
FFLO phase~\cite{fflo}, can appear in the presence of external magnetic 
field or intrinsic ferromagnetic order, and could thereby permit a larger 
threshold of the spin exchange energy to coexist with superconductivity.  

On the other hand, it has been known since the early days of research on 
$^3$He that alternative superconducting states, other than $s$-wave, can 
be favoured in a ferromagnetic background. The early theories of an
equivalent phenomenon to occur in the solid-state have been formulated
in the early 1980s~\cite{fay-appel}, despite the absence of any
experimental example of a ferromagnetic superconductor at the
time \cite{footnote-ZrZn2}. With the discovery of superconductivity in
UGe$_2$ and ZrZn$_2$, especially given that the same electrons
are believed to participate both in ferromagnetism and SC, this
latter scenario had to be taken seriously to explain the microscopic
coexistence between the two phases. In particular, the very large hyperfine magnetic molecular
field in these materials, measured~\cite{tsutsui-muSR} \eg with M{\"o}ssbauer spectroscopy,
far exceeds the Pauli limit. This excludes any possibility of
singlet-pairing superconductivity. 

We should note that although the latter statement is true in UGe$_2$ and other ferromagnetic 
superconductors, one may still ask whether in principle a singlet-type superconductivity can 
coexist with ferromagnetism. Although some early theoretical studies~\cite{bedell} indicated 
that the answer to this question may be affirmative provided FM is weak, a more careful analysis 
concluded~\cite{breakdown} that the coexistence state of spin-singlet pairing and ferromagnetism 
always turns out to be energetically unfavorable against the non-magnetic superconducting state
even if a finite-momentum pairing (FFLO) state is considered. Later, it was proposed \cite{cuoco} 
that the coexistence of metallic ferromagnetism and singlet superconductivity may be realized
assuming that the magnetic instability is due to kinetic exchange. However, the coexistence of 
magnetism and spin-triplet superconductivity appears to be a more promising scenario, since the 
Cooper pairs may use their spin degree of freedom to align themselves with the internal magnetic 
field.

An  experimental fact that is even more striking is that in
all ferromagnetic superconductors  known to date, the SC phase is only observed in a small part of
the phase diagram otherwise occupied by ferromagnetism~\cite{footnote-URhGe}, and it is the
region where the magnetism appears to be at its weakest that SC sets in -- on the boundary with paramagnetism 
when the Curie temperature is driven to zero (typically by applying pressure).
This immediately raises the question of the microscopic origin of SC pairing, and 
whether ferromagnetic spin fluctuations play the role of a ``glue'' for Cooper pair formation very much 
as they do in superfluid $^3$He.  It is equally interesting what role the zero-temperature pressure-tuned phase transition 
plays in formation of superconductivity and whether notions involving quantum criticality 
(provided the phase transitions are second order) are necessary to explain the phenomenon. 

Although there is no universal answer to this question yet and the research efforts, both 
experimental and theoretical, are focused on this issue, it is interesting to note that in both UGe$_2$ and ZrZn$_2$ 
the ferromagnetic phase transition as a function of pressure becomes 1st order as the ``critical pressure'' 
is approached at $T=0$. One cannot therefore straightforwardly apply a theory of quantum criticality 
(be it the Hertz-Millis~\cite{hertz-millis} theory or one of its variations) given the absence of the
quantum critical point as such. It is undeniable however that the point where Curie temperature goes to zero is
of crucial importance to the formation of the SC state.

Drawing further parallels between triplet-pairing FM superconductors and the superfluid $^3$He, 
one may wonder whether different symmetries of the SC gap can occur, as is the case in the different 
phases~\cite{leggett} of $^3$He. For example, can the gap symmetry with point or line nodes be 
realized in the ferromagnetic superconductors?
Very recently, experimental evidence has appeared which suggests that the answer is `yes'. 
Harada~\etal~\cite{harada} reported on $^{73}$Ge nuclear-quadrupole-resonance 
experiments performed under pressure, in which the nuclear spin-lattice relaxation rate revealed an 
unconventional nature of superconductivity implying that the majority spin band in UGe$_2$ was gapped with line nodes,
while the minority spin band remained gapless at the Fermi level.

Motivated by this, we present  a mean-field model for coexisting ferromagnetism and spin-triplet superconductivity 
with a SC order parameter that displays line nodes in majority-spin channel and is gapless for minority spin. 
We first study the interplay between the magnetic and superconducting order parameters, and then proceed to make 
several predictions for experimentally relevant quantities: the specific heat capacity, 
Knight shift, and tunneling conductance. Let us briefly summarize our main results. We find that the low-temperature specific heat capacity $C_V$ shows power-lawer behaviour (to be contrasted with the conventional exponential decay in the $s$-wave case), and that the gapless minority spins dominate the contribution to $C_V$ at low temperatures, giving rise to a linear $T$-dependence. Also, the relative jump in $C_V$ shows a strong dependence on the exchange splitting in the system. With regard to the Knight shift, we find that it is suppressed at $T=0$ with increasing exchange splitting of the majority and minority spin bands when the external field is applied perpendicular to the spin of the Cooper pairs in the system. In general, however, it depends strongly on the orientation of the field with respect to the crystallographic axes of the compound, indicative of the triplet pairing in the system. Finally, the normalized tunneling conductance spectra show a strong directional dependence with respect to the orientation of the superconducting order parameter in reciprocal space, but change very little upon modifying the exchange splitting in the system. Our findings should be useful for comparison with experimental studies, and could lead to further insights as regards the nature of the superconducting order parameter. 

\indent 
This paper is organized as follows. We first describe the phenomenological
framework to be used in this work in Sec. \ref{sec:review}. 
We then present our theoretical model in Sec. \ref{sec:theory}, and provide the results of the self-consistent mean-field treatment at both zero and finite temperatures in Sec. \ref{sec:results1}. We then proceed to make predictions for experimentally accessible quantities in Sec. \ref{sec:results2}, using the self-consistently obtained results from Sec. \ref{sec:results1}. We discuss our findings in Sec. \ref{sec:discuss}, and summarize in Sec. \ref{sec:summary}. We will use boldface notation for vectors, $\hat{...}$ for operators,  $\check{...}$ 
for 2$\times$2 matrices, and $\hat{...}$ for 4$\times$4 matrices.

\section{Phenomenological framework}\label{sec:review}

The issue of coexisting ferromagnetism and superconductivity dates back to half a century ago when the celebrated 
FFLO state was  predicted~\cite{fflo} as a finite-momentum pairing state 
with real-space structure of the singlet SC order 
parameter that may develop under certain conditions close to the critical magnetic field $H_{c2}$. 
The conditions for the FFLO state are however very different from those observed in ferromagnetic superconductors such
as UGe$_2$. In particular, as has already been emphasised above, the magnetic molecular field felt by Cooper pairs inside
the ferromagnet is many times larger\cite{tsutsui-muSR} than the Pauli limiting field necessary to destroy 
the singlet Cooper pairs. We shall therefore concentrate on triplet-type superconducting pairing. 

Several remarks are in order.
 We note from the outset that the ferromagnetism observed in the Uranium compounds and in ZrZn$_2$ is itinerant, 
Stoner-like in its nature. We shall therefore not discuss the topic of localized magnetic moments that would have, 
among other things, provided a pair-breaking
mechanism in accord with Abrikosov--Gor'kov theory~\cite{abrikosov-gorkov} of magnetic scattering.
Here, we will assume that the same electrons involved in the spontaneous SU(2) symmetry breaking associated with 
ferromagnetism, also participate in the U(1) gauge symmetry breaking that characterizes a superconductor.

The idea of triplet pairing occurring between the same electrons that form the Stoner instability at the border of ferromagnetism
goes back to Fay and Appel\cite{fay-appel} (1980) who considered exchange of magnetic spin fluctuations as a microscopic
mechanism for Cooper pairing. 
More recently, the problem has been revisited~\cite{ohmi, machida, walker, nevidomskyy, linder} in the light of experimental 
findings in UGe$_2$ and other ferromagnetic superconductors.

In this paper, we shall take a phenomenological approach to superconductivity, leaving the intriguing and debated 
question of the microscopic mechanism for Cooper pairing aside. In particular,  we shall consider systems where
superconductivity appears at a lower temperature than the temperature at which onset of 
ferromagnetism is found. This is certainly the case experimentally and may be simply due to the fact that the energy 
scales for the two phenomena are quite different, with the exchange energy naturally being the largest. 
It may, however, also be due to the fact that superconductivity is dependent on ferromagnetism for its very existence. 
Such a suggestion has recently been put forth \cite{niu}. 

A crucial issue to address in this context is whether superconductivity and ferromagnetism
are phase-separated (such as \eg solid and liquid phases coexisting at 
the melting point) or not. Fairly strong experimental evidence for non-phase-separated coexistence of 
ferromagnetism and superconductivity has recently been presented in UGe$_2$ ~\cite{kotegawa2005}. 
However, even if such non-phase-separated coexistence is established, there still remains
the issue of whether the superconducting order-parameters exhibits spatial variations, precisely  
due to its non-phase-separated coexistence with ferromagnetic order. One obvious candidate for such spatial 
variations \cite{tewari2004} is a spontaneously formed Abrikosov vortex lattice, induced by the internal 
magnetization $\mathbf{M}$. As argued in Ref.~\onlinecite{mineev2005}, an important factor with respect 
to whether a vortex lattice appears or not could be the magnitude of the internal magnetization $\mathbf{M}$. 
Specifically, Ref.~\onlinecite{mineev1999} suggested that vortices may arise if $4\pi\mathbf{M}>\mathbf{H}_{c1}$, 
where $\mathbf{H}_{c1}$ is the lower critical field. It is conceivable that a weak ferromagnetic state coexisting 
with superconductivity may give rise to a domain structure, in the absence of an external field, that is 
vortex-free. Therefore, we shall consider non-phase-separated coexistence of the 
FM and  SC order parameters from here on, as have other studies \cite{shopova2005}. We will also leave the 
complications arising from the spatial variation of the superconducting order parameter originating with a putative spontaneously formed Abrikosov vortex lattice in the superconducting 
order parameter for future investigations.

Spin-triplet superconductors are characterized by a multicomponent order parameter, 
which for the simplest case of the $p$-wave may be expressed in terms of three independent 
components of a ${\bf d}$-vector:
\begin{equation}
\mathbf{d}_\vk 
= \Big[\frac{\Delta_{\vk\downarrow\downarrow} - \Delta_{\vk\uparrow\uparrow}}{2}, \frac{-\i(\Delta_{\vk\downarrow \downarrow}+\Delta_{\vk\uparrow\uparrow})}{2}, \Delta_{\vk\uparrow\downarrow}\Big].
\label{d_k}
\end{equation}
Note that $\mathbf{d}_\vk$ transforms like a vector under spin rotations. In terms of the 
components of  $\mathbf{d}_\vk$, the order parameter itself is a 2x2 matrix that reads
\begin{equation}
\check{\Delta}_{\alpha\beta}(\kk)\equiv \langle c_{\kk,\alpha} c_{\kk,\beta} \rangle = 
[ i(\mathbf{d}_\kk\cdot\boldsymbol{\sigma}) \sigma_y ]_{\alpha\beta},
\end{equation}
where $\boldsymbol{\sigma}$ is the vector of Pauli matrices, and $c^\dag_{\kk,\alpha}, c_{\kk,\alpha}$ are the usual 
electron creation-annihilation operators for momentum $\vk$ and spin $\alpha$. 

The superconducting order parameter is characterized as \emph{unitary} if the modulus of the gap is proportional 
to the unity matrix: $(\check{\Delta}\cdot\check{\Delta}^\dagger) \propto \check{1}$. Written in terms of the vector  
$\dd_\kk$, this condition is equivalent to the requirement that $\langle \mathbf{S}_\kk \rangle=0$, where we have 
introduced the net magnetic moment (or spin) of the Cooper pair
\be
 \langle \mathbf{S}_\kk \rangle \equiv \i(\mathbf{d}_\vk\times\mathbf{d}_\vk^*).
\ee
The \emph{unitary} triplet state thus has Cooper pairs with zero magnetic moment, whereas the 
\emph{non-unitary} state is characterized by non-zero value of  $\langle \mathbf{S}_\kk \rangle\neq 0$.
The latter effectively means that time-reversal symmetry is spontaneously broken in the spin part of the 
Cooper pairs~\cite{footnote-orbital}. It is thus intuitively clear that having the spin of the Cooper pair 
aligned with the internal magnetic field of the ferromagnet can lower the energy of the resulting coexistence 
state. The above argument that the order parameter in the ferromagnetic superconductors must be non-unitary 
has been put forward by Machida and Ohmi\cite{machida}, and others\cite{hardy2005,walker}. Distinguishing 
between unitary and non-unitary states in ferromagnetic superconductors is clearly one of the primary 
objectives in terms of identifying the correct SC order parameter. To this end, recent studies have 
focused on calculating transport properties of ferromagnetic superconductors
~\cite{linderPRL,linderPRB,linder_gronsleth,yokoyama,brataas,shen}. There have also been investigations of identifying spin-triplet pairing in quasi-1D materials \cite{lebed,yacovenko,giamarchi,sademelo}.

Finally, we note that inter-subband pairing is expected to be strongly suppressed in the presence of the 
Zeeman splitting between the $\uparrow,\downarrow$ conduction sub-bands. In other words, only electrons within the same 
subband will form Cooper pairs (the so-called \emph{equal-spin pairing}) and  we shall set $\Delta_{\uparrow\downarrow}=0$ 
in what follows.
Moreover, the requirement of non-unitarity of the order parameter then reduces  to the requirement that
the vector $\mathbf{d}_\kk$ in Eq.~(\ref{d_k}) should have two non-zero components, 
i. e. $\Delta_{\uparrow\uparrow} \neq \Delta_{\downarrow\downarrow}$, which one would expect anyway in the presence of
the Zeeman splitting between the two spin subbands.  The spin of the Cooper pair is then 
$\langle S_z \rangle = \frac{1}{2}(|\Delta_{\uparrow\uparrow}|^2 - |\Delta_{\downarrow\downarrow}|^2)$ and is aligned 
along the magnetic field ($z$ being the spin quantization axis).

\section{Theory}\label{sec:theory}
We consider a model of a ferromagnetic superconductor described by uniformly coexisting itinerant ferromagnetism and non-unitary, spin-triplet superconductivity. We write down a weak-coupling mean-field theory Hamiltonian with equal-spin pairing Cooper pairs and a finite magnetization along the easy-axis similar to the model studied in Refs.~\onlinecite{nevidomskyy, linder}, namely
\begin{align}
\hat{H} &= \sum_\vk \xi_\vk + \frac{INM^2}{2} - \frac{1}{2}\sum_{\vk\sigma} \Delta_{\vk\sigma\sigma}^\dag b_{\vk\sigma\sigma} \notag\\
&+\frac{1}{2}\sum_{\vk\sigma} \Big(\hat{c}_{\vk\sigma}^\dag \hat{c}_{-\vk\sigma}\Big)
\begin{pmatrix}
\xi_{\vk\sigma} & \Delta_{\vk\sigma\sigma} \\
\Delta_{\vk\sigma\sigma}^\dag & -\xi_{\vk\sigma} \\
\end{pmatrix}
\begin{pmatrix}
\cop_{\vk\sigma}\\
\cop_{-\vk\sigma}^\dag\\
\end{pmatrix},
\end{align}
where $b_{\vk\sigma\sigma} = \langle c_{-\vk\sigma} c_{\vk\sigma}\rangle$ is the non-zero expectation value of the pair of Bloch states.
Applying a standard diagonalization procedure, we arrive at 
\begin{align}
\hat{H} &= H_0 + \sum_{\vk\sigma} E_{\vk\sigma} \hat{\gamma}_{\vk\sigma}^\dag\hat{\gamma}_{\vk\sigma},\notag\\
H_0 &= \frac{1}{2}\sum_{\vk\sigma}(\xi_{\vk\sigma} - E_{\vk\sigma} - \Delta_{\vk\sigma\sigma}^\dag b_{\vk\sigma\sigma}) + \frac{INM^2}{2},
\end{align}
where $\{\hat{\gamma}_{\vk\sigma},\hat{\gamma}_{\vk\sigma}^\dag\}$ are new fermion operators and the eigenvalues read
\begin{equation}
E_{\vk\sigma} = \sqrt{\xi_{\vk\sigma}^2 + |\Delta_{\vk\sigma\sigma}|^2}.
\end{equation}
It is implicit in our notation that $\xi_\vk = \varepsilon_\vk - E_F$ is measured from the Fermi level, where $\varepsilon_\vk$ is the kinetic energy. The free energy is obtained through
\begin{align}
F = H_0 - \frac{1}{\beta}\sum_{\vk\sigma}\text{ln}(1 + \e{-\beta E_{\vk\sigma}}),
\end{align}
such that the gap equations for the magnetic and superconducting order parameters become \cite{nevidomskyy}
\begin{align}\label{eq:mfequations}
M &= -\frac{1}{N} \sum_{\vk\sigma} \frac{\sigma \xi_{\vk\sigma}}{2 E_{\vk\sigma}} \text{tanh}(\beta E_{\vk\sigma}/2),\notag\\
\Delta_{\vk\sigma\sigma} &= -\frac{1}{N}\sum_{\vk'} V_{\vk\vk'\sigma\sigma} \frac{\Delta_{\vk'\sigma\sigma}}{2 E_{\vk'\sigma}}\text{tanh}(\beta E_{\vk'\sigma}/2).
\end{align}
Specifically, we now consider a model which should be of relevance to the
ferromagnetic superconductor UGe$_2$, and possibly also for UCoGe and
URhGe. In Ref. ~\onlinecite{harada}, it was argued that the majority
spin (spin-up in our notations) fermions were gapped and that the order parameter displayed line
nodes, while the minority (spin-down) fermions remained gapless at the Fermi level in
the heavy-fermion compound UGe$_2$. An obvious mechanism for
suppressing the superconducting instability in the minority-spin
channel as compared to the majority-spin channel is the difference in
density of states (DOS) at the Fermi level. Indeed, from Fig. 1
in Ref.~\onlinecite{nevidomskyy} (see also Fig. 4 in
Ref.~\onlinecite{linder}), it is seen that the critical temperature
for pairing in the minority-spin subband, $T_c^\downarrow$, is predicted
to be much smaller than the critical $T_c^\uparrow$ for the majority-spin subband, even for quite weak
magnetic exchange splittings. 
Given the already quite low critical temperature $T_c$
that is observed experimentally in ferromagnetic superconductors
($T_c\lesssim 1$~K), which we associate with $T_c^\uparrow$, we
therefore conclude that it might indeed be very hard to observe
experimentally the even smaller gap in the minority-spin subband. 
Therefore,  it is permissible to only consider pairing in the
majority-spin channel and neglect a small (if any) pairing between
minority-spin electrons.
In our notation this means setting $M\neq0, \Delta_{\vk\uparrow}\neq0, \Delta_{\vk\downarrow}=0$. 
\par
We stress that the above statement, although intuitively attractive, may need further
justification since we have so far neglected completely the spin-orbit
interaction that is expected to be strong in Urainium based compounds,
such as UGe$_2$, URhGe and UCoGe. The effect of the latter would be to
provide some effective coupling between majority and minority spin
subbands and would probably lead to induced SC order parameter in
minority spin channel. This issue is left for future
study~\cite{soc-future}.

To model the presence of line nodes in the order parameter, we choose
\begin{equation}\label{eq:gaptriplet}
\Delta_{\vk\uparrow} =  \Delta_{\bar{\vk}_F\uparrow\uparrow} = \Delta_0\cos\theta,
\end{equation}
where $\bar{\vk}_F$ is the normalized Fermi wave-vector, such that the gap only depends on the direction of the latter. This is the weak-coupling approximation. The above gap satisfies the correct symmetry requirement dictated by the Pauli principle, namely a sign change under inversion of momentum, $\theta \to \pi-\theta$. Here, $\theta$ is the azimuthal angle in the $xy$-plane. Our choice of this particular symmetry for the $p$-wave superconducting gap is motivated by the experimental results of Harada \etal \cite{harada}. The $\cos\theta$-dependence is also in accord with the results of Ref.~\onlinecite{shick}, which showed that the majority band at the Fermi level for UGe$_2$ is strongly anisotropic with a small dispersion along the $k_y$-direction. We consider here a situation where the electrons are restricted from moving along the $\mathbf{z}$-axis. The motivation for this is that, strictly speaking, it seems plausible that uniform coexistence of ferromagnetic and superconducting order should only be realized in thin-film structures where the Meissner (diamagnetic) response of the superconductor is suppressed for in-plane magnetic fields. The thin-film structure would then also suppress the orbital effect of the field. In a bulk structure, as considered in Ref.~\onlinecite{bedell}, we expect that a spontaneous vortex lattice should be the favored thermodynamical state \cite{tewari2004}, unless prohibited by a possible domain structure. Having said that, we point out that there is no firm
experimental evidence for the presence of such a vortex phase in
ferromagnetic superconductors such as UGe$_2$ and ZrZn$_2$, and we therefore do
not exclude some mechanism that would instead stabilise a truly uniform
coexistence of the SC and FM in these materials.
It should be mentioned that uniform coexistence of ferromagnetism and superconducting order have also been speculated to occur in quasi-1D and quasi-2D materials such as RuSr$_2$GdCu$_2$O$_8$. \cite{ru} In our model, the pairing potential may be written as
\begin{align}
V(\theta,\theta') = -g\cos\theta\cos\theta',
\end{align}
where $g$ is the weak-coupling constant. Conversion to integral equations is accomplished by means of the identity
\begin{equation}
\frac{1}{N} \sum_\vk f(\xi_{\vk\sigma}) = \int \text{d}\varepsilon N^\sigma(\varepsilon),
\end{equation}
where $N^\sigma(\varepsilon)$ is the spin-resolved density of states. In three spatial dimensions, this may be calculated from the dispersion relation by using the formula
\begin{equation}
N^\sigma(\varepsilon) = \frac{V}{(2\pi)^3} \int_{\varepsilon_{\vk\sigma} = \text{const}} \frac{\text{d} S_{\varepsilon_{\vk\sigma}}}{|\hat{\nabla}_\vk \varepsilon_{\vk\sigma}|}.
\end{equation}
With the dispersion relation $\xi_{\vk\sigma}= \varepsilon_\vk - \sigma IM - E_F$, one obtains
\begin{equation}
N^\sigma(\varepsilon) = \frac{mV\sqrt{2m(\varepsilon + \sigma IM + E_F)}}{2\pi^2}.
\end{equation}
In their integral form, Eqs. (\ref{eq:mfequations}) for the order parameters read
\begin{align}\label{eq:gapeqint}
M &= -\frac{1}{4\pi}\sum_\sigma \sigma \int^{2\pi}_0 \int_{-E_F-\sigma IM}^{\infty} \text{d}\theta \text{d}\varepsilon \frac{\varepsilon N^\sigma(\varepsilon)}{E_\sigma(\varepsilon,\theta)}\notag\\
&\hspace{0.3in} \times\text{tanh}[\beta E_\sigma(\varepsilon,\theta)/2],\notag\\
1 &= \frac{g}{4\pi} \int^{2\pi}_0\int^{\omega_0}_{-\omega_0} \text{d}\theta\text{d}\varepsilon \frac{ N^\uparrow(\varepsilon)\cos^2\theta }{E_\uparrow(\varepsilon)}\text{tanh}[\beta E_\uparrow(\varepsilon,\theta)/2].
\end{align}
For ease of notation, we also define
\begin{align}
\Delta_\sigma(\theta) &=\left\{ \begin{array}{ll}
 \Delta_0\cos\theta &\text{if } \sigma=\uparrow\\
0&\text{if } \sigma=\downarrow \\
\end{array} \right\},\notag\\
E_\sigma(\varepsilon,\theta) &=\left\{ \begin{array}{ll}
 \sqrt{\varepsilon^2 + \Delta_0^2\cos^2\theta}. &\text{if } \sigma=\uparrow\\
\varepsilon &\text{if } \sigma=\downarrow \\
\end{array} \right\}.
\end{align}
For the following treatment, we define $\tilde{M} = IM/E_F$, \ie the exchange energy scaled on the Fermi energy. Moreover, we set $c = gN(0)/2$ 
to a typical value of 0.2 and $\tilde{\omega}_0 = \omega_0/E_F = 0.01$ as the typical spectral width of the bosons responsible for the 
attractive pairing potential. Finally, we define the parameter $\tilde{I}=IN(0)$ as a measure of the magnetic exchange coupling. As 
discussed below, only for $\tilde{I}>1$ will a spontaneous magnetization appear in our model, in agreement with the Stoner criterion
for itinerant ferromagnetism.

\section{Results: mean-field model for coexistence}

\subsection{Zero temperature case}\label{sec:results1}

For zero-temperature, the superconducting gap equation reads 
\begin{align}\label{eq:gap}
1 = \frac{g}{4\pi}\int^{2\pi}_0 \int^{\omega_0}_{-\omega_0} \text{d}\theta\text{d}\varepsilon \frac{N^\uparrow(\varepsilon)\cos^2\theta}{\sqrt{\varepsilon^2+\Delta_0^2\cos^2\theta}}.
\end{align}
Under the assumption that $\omega_0\gg\Delta_0$, we obtain that
\begin{align}
\frac{2}{c\sqrt{1+\tilde{M}}} = \text{ln}\Big(\frac{2\omega_0}{\Delta_0}\Big) - \frac{1}{\pi}\int^{2\pi}_0 \text{d}\theta\cos^2\theta \text{ln}|\cos\theta|.
\end{align}
which may be solved to yield the zero-temperature gap
\begin{align}\label{eq:gap1}
\Delta_0 = 2.426\omega_0\exp[-2/(c\sqrt{1+\tilde{M}})].
\end{align}
By inserting Eq. (\ref{eq:gap1}) into the gap equation for the magnetization in Eq. (\ref{eq:gapeqint}), we have managed to decouple 
the self-consistency equations for $M$ and $\Delta_0$. Numerical evaluation reveals that the gap equation for $M$ is completely 
unaffected by the presence of $\Delta_0$, which physically means that the magnetization remains unaltered with the onset of 
superconductivity. This is reasonable in a model where the energy scale for the onset of magnetism
is vastly different from the energy scale for superconductivity, such that by the time superconductivity
sets in, the ordering of the spins essentially exhausts the maximum possible magnetisation.  
\par
\begin{figure}[h!]
\centering
\resizebox{0.45\textwidth}{!}{
\includegraphics{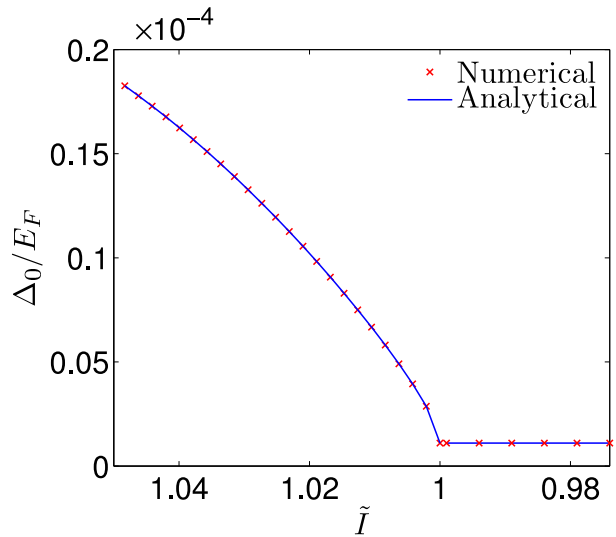}}
\caption{(Color online) The gap-dependence on the ferromagnetic exchange interaction parameter $\tilde{I}=IN(0)$. The gap remains constant for $\tilde{I}\in[0,1]$, corresponding to a unitary phase. The gap $\Delta_0$ then starts growing with increasing $\tilde{I}$ for $\tilde{I}>1.0$, announcing the onset of a spontaneous magnetization. The analytical formula is based on Eq. (\ref{eq:gap1}). }
\label{fig:GapT0}
\end{figure}

The dependence of $\Delta_0$ on $\tilde{I}$ is shown in Fig. \ref{fig:GapT0}. The gap remains constant for $\tilde{I}\in[0,1]$, 
which is a unitary phase. In the unitary phase, there is no reason for the minority spin band to remain ungapped when 
$M=0$, and hence we would expect two gaps $\Delta_\uparrow=\Delta_\downarrow$ of equal magnitude for $\tilde{I}<1$. Our model 
of gapping exclusively for the majority spin band is therefore justified only for $\tilde{I}>1$, which is the regime we shall 
be concerned with throughout this article. The onset of a spontaneous magnetization for $\tilde{I}>1$ is the well-known Stoner 
criterion for an isotropic electron gas, where the spin susceptibility may be written 
as \cite{sigrist}
\begin{align}
\chi(\mathbf{q},\omega) &= \frac{\chi_0(\mathbf{q},\omega)}{1 - I\chi_0(\mathbf{q},\omega)},\notag\\
\chi_0(\mathbf{q},\omega) &= N_0\Big( 1 - \frac{\mathbf{q}^2}{12k_F^2} + \i\frac{\pi\omega}{2v_F|\mathbf{q}|}\Big),\notag\\
&|\mathbf{q}| \ll 2k_F,\;\; \omega\ll E_F.
\end{align}
For a parabolic band, the static susceptibility is maximal for $\mathbf{q}=0$ where 
\begin{equation}
\chi(\mathbf{q}=0, \omega=0) = \frac{N_0}{1-IN(0)} = \frac{N_0}{1-\tilde{I}}.
\end{equation}
The introduction of a ferromagnetic order is demarcated by the divergence of the susceptibility for $\tilde{I}=1$, which is 
precisely Stoner's criterion for itinerant ferromagnetism. In the absence of superconductivity, the 
self-consistency equation for the 
magnetization at $T=0$ reduces to
\begin{align}\label{eq:mag0}
h = -\frac{\tilde{I}}{3\sqrt{E_F}}\sum_\sigma \sigma[(E_F+\sigma h + \eta)^{3/2} - 2(E_F+\sigma h)^{3/2}],
\end{align}
where $\eta$ is an upper energy cut-off determined by the band-width and $h=IM$ is the exchange splitting of the majority and minority bands. Since the energy scales for the magnetic and superconducting order parameter differ so greatly in magnitude, Eq. (\ref{eq:mag0}) is an excellent approximation even in the coexistent state (we have verified this numerically). 

\subsection{Finite temperature case}\label{sec:results2}

\begin{figure}[h!]
\centering
\resizebox{0.45\textwidth}{!}{
\includegraphics{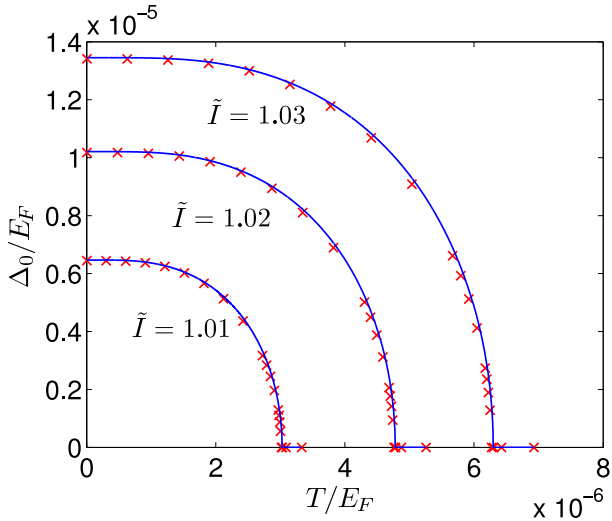}}
\caption{(Color online) Self-consistently obtained solution for the superconducting gap $\Delta_0$ (red symbols) compared to the analytical expression Eq. (\ref{eq:temp}) with $\gamma=1.70$ (blue lines), modelling a BCS-like temperature dependence.}
\label{fig:GapT}
\end{figure}

\begin{figure}[h!]
\centering
\resizebox{0.45\textwidth}{!}{
\includegraphics{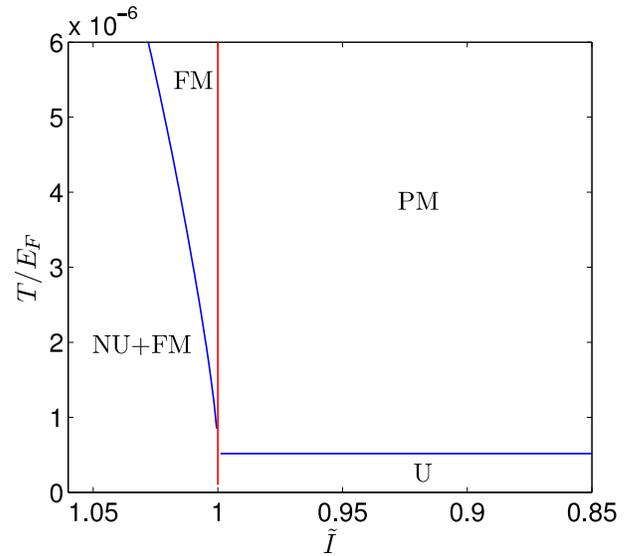}}
\caption{(Color online) The phase-diagram of our model in the $T$-$\tilde{I}$ plane. For $\tilde{I}>1.0$, a spontaneous magnetization arises and allows for the possible uniform coexistence of ferromagnetism and triplet superconductivity. Note that decreasing $\tilde{I}$ (going from left to right along the x-axis) corresponds to an increasing external pressure $p$. The abbrevations stand for non-unitary (NU), unitary (U), ferromagnetic (FM), and paramagnetic (PM).}
\label{fig:phase}
\end{figure}

\noindent The critical temperature for the superconducting order parameter is obtained in the standard way [setting $\Delta_0=0$ in Eq. (\ref{eq:gap})] to yield
\begin{equation}\label{eq:Tc}
T_c = 1.134\omega_0\exp[-2/(c\sqrt{1+\tilde{M}})].
\end{equation}
In Fig. \ref{fig:GapT}, we plot the temperature-dependence of the self-consistently obtained solution of $\Delta_0$ and compare it to the analytical mean-field temperature dependence
\begin{equation}\label{eq:temp}
\Delta_0(T) = \Delta_0(0)\text{tanh}[\gamma\sqrt{T_c/T - 1}].
\end{equation}
The BCS result is $\gamma=1.74$, but we find a better fit for our numerical results using $\gamma=1.70$. Throughout the rest of this paper, we shall therefore make use of Eq. (\ref{eq:temp}) with $\gamma=1.70$ to model the temperature-dependence of the gap for $\tilde{I}=\{1.01,1.02,1.03\}$, since the agreement is excellent with the full  numerical solution. As in the zero-temperature case, we find that the gap equations in Eq. (\ref{eq:gapeqint}) may be completely decoupled also at finite temperature. We have verified that the gap equation for the superconducting order parameter has a unique non-trivial solution, which guarantees that the system will prefer to be in the coexistent state of ferromagnetism and superconductivity.
The phase-diagram of the model we are considering may be obtained numerically and is shown in Fig. \ref{fig:phase}. As seen, a quantum 
phase transition may occur at $\tilde{I}=1.0$, separating the 'unitary' superconducting state (see discussion in an earlier paragraph) 
from the ferromagnetic, non-unitary superconducting state. The critical temperature for the magnetic order parameter is 
orders of magnitudes larger than $T_c$ for the superconductivity except for very close to $\tilde{I}=1.0$.  The increase in $T_c$ in the non-unitary phase as compared to the unitary phase is a result of the increase in density of states with magnetization for the majority spin.
\par
Experimentally, one often maps out the $T$-$p$ phase diagram, where $T$ is temperature and $p$ is pressure. Note that the value of 
$\tilde{I}$ may be controlled experimentally by adjusting the pressure on the sample. A change in pressure is accompanied by a 
change in the width of the electron bands, and therefore directly affects the density of states at the Fermi level: increasing the 
pressure on the samples reduces the density of states, and hence also the effective coupling constant $\tilde{I}$.\cite{wolfarth}. 
A notable feature in the phase diagram for UGe$_2$ as determined experimentally, is that superconductivity only appears in the 
ferromagnetic phase, and not in the paramagnetic phase. 

\section{Results: experimental predictions}
\begin{figure}[h!]
\centering
\resizebox{0.45\textwidth}{!}{
\includegraphics{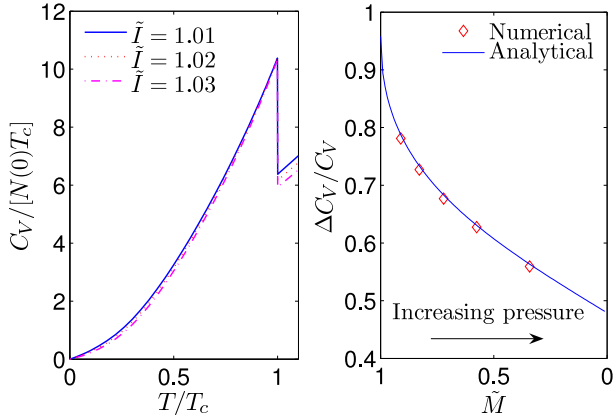}}
\caption{(Color online) The left panel shows a plot of the specific heat capacity, using self-consistently obtained order parameters, for 
three different values of $\tilde{I}$. The right panel shows relative jump (superconducting vs. normal state) of the specific heat at the 
transition temperature as a function of the normalized exchange splitting between the spin bands, $\tilde{M}$. Numerically calculated 
values are shown in red, analytical result [Eq. (\ref{eq:reljump})] using $\gamma\approx1.70$ are shown in blue. }
\label{fig:heat}
\end{figure}

We next proceed to using the self-consistently obtained solutions from the previous section to make predictions for three experimental 
quantities that are routinely used to study superconducting condensates: specific heat, Knight shift, and tunneling conductance spectra. We first consider the normalized heat capacity, which is defined 
as
\begin{align}
C_V &= \frac{\beta^2}{8\pi} \sum_\sigma \int^{2\pi}_0 \int^\infty_{-E_F-\sigma IM}\text{d}\theta\text{d}\varepsilon \frac{N^\sigma(\varepsilon)}{\cosh^{2}[\beta E_\sigma(\varepsilon,\theta)/2]}\notag\\
&\times \Big[ E_\sigma^2(\varepsilon,\theta) - T\Big(\Delta_\sigma(\theta) \frac{\partial \Delta_\sigma(\theta)}{\partial T} 
- \sigma\varepsilon I \frac{\partial M}{\partial T}  \Big) \Big].
\end{align}
Since the critical temperature of $M$ is much larger than the critical temperature for $\Delta_0$ in our model, we may safely 
neglect $\partial M/\partial T$ in the low-temperature regime. Consider Fig. \ref{fig:heat} for a plot of the specific heat 
capacity using three representative values for $\tilde{I}$. The general trend with increasing $\tilde{I}$ is an increase of 
the jump of $C_V$ at $T=T_c$. The physical reason for this is that the majority spin carriers will dominate the jump in specific 
heat stronger when the exchange splitting between the bands increases, which is in agreement with the results of 
Ref.~\onlinecite{linder}. Analytically, the relative jump in specific heat may be expressed as
\begin{align}\label{eq:reljump}
\Big( \frac{\Delta C_V}{C_V}\Big) \Big|_{T=T_c} \sim \Bigg( 1 + \sqrt{\frac{1-h/E_F}{1+h/E_F}}\Bigg)^{-1}.
\end{align}
It depends on the exchange splitting in the superconductor since the contribution from the majority spin carriers will tend to dominate the specific heat when $h$ increases.
The low-temperature scaling with $T$ bears witness of the line nodes in the gap, and is to 
be contrasted with the more rapidly decaying $s$-wave case. Also note that the minority spin fermions are in the normal state and give a significant contribution to the specific heat in form of a linear $T$-dependence at low temperatures. If both spin species were gapped with line nodes, one would expect a $T^2$-dependence of the low temperature specific heat.
\par
In the experimental study of the heat-capacity in UGe$_2$ conducted in Ref.~\onlinecite{tateiwa}, a peak of the heat-capacity 
associated with the superconducting transition was
observed in a narrow pressure region $\Delta p \simeq 0.1$ GPa around $p_x$. Here, $p_x$ is the pressure at which the superconducting transition
temperature $T_c$ shows a maximum value. Farther away from $p_x$, the heat capacity anomaly was smeared out. In particular, a substantial 
residual value of $C_V/T$ was observed at $T\to0$. The authors of Ref.~\onlinecite{tateiwa} argued that neither the minority band density 
of states at the Fermi level nor the contribution from a self-induced vortex state would be appropriate to describe this residual value. Instead, 
it might stem from impurities that induce a finite density of states at the Fermi level. For an anisotropic superconductor like UGe$_2$, the 
residual value would be highly sensitive to such impurities. It is also clear that the observation of sharp peaks, similar to the ones we 
obtain in Fig. \ref{fig:heat}, depend strongly on the applied pressure on the superconductor, and in particular to how close it is to $p_x$.
\par
We next consider the spin susceptibility, making use of the standard formula \cite{schrieffer}
\begin{equation}\label{eq:spin}
\chi(\mathbf{q},\omega) = -\frac{1}{2\beta}\sum_{\vk,\i\omega_n} \text{Tr}\{\hat{\mathcal{G}}(\vk,\i\omega_n)\hat{\mathcal{G}}(\vk+\mathbf{q},\i\omega+\i\omega_n)\},
\end{equation}
where $\hat{\mathcal{G}}$ is the matrix Green's function in particle-hole and spin-space, where $\omega_n=2(n+1)\pi/\beta$ 
are fermionic Matsubara frequencies. In the static ($\omega=0$) and uniform ($\mathbf{q}=0$) limit, Eq. (\ref{eq:spin}) 
reduces to the Knight shift $\kappa \equiv \chi(0,0)$. We define the normalized Knight shift as 
\begin{equation}\label{eq:knight}
\frac{\kappa}{\kappa_0} 
= \frac{\beta}{8\pi} \sum_\sigma \int^{2\pi}_0 \int^{\infty}_{-E_F - \sigma IM} 
\frac{\text{d}\theta\text{d}\varepsilon N_\sigma(\varepsilon)}{\cosh^2[\beta E_\sigma(\varepsilon,\theta)/2]}.
\end{equation}
The Knight shift is a measure of the polarizibility of the conduction electrons in the compound, and serves as a highly useful probe to 
distinguish between singlet and triplet superconductivity. For a singlet superconductor, the total spin $S$ of the Cooper pair is zero, 
and the Knight shift therefore vanishes at $T=0$ since there are no quasiparticle excitations in the superconductor that may be polarized. 
The Knight shift vanishes regardless of the direction in which the external magnetic is applied for a singlet superconductor. For a 
triplet superconductor, this is quite different. The Knight shift now may be anisotropic in terms of the direction in which the 
magnetic field is applied. By means of the $\mathbf{d}_\vk$-vector formalism [see Eq. (\ref{d_k})], one may infer that the Knight shift is unaltered even for $T<T_c$ when 
$\mathbf{d}_\vk \perp \mathbf{H}$, but is altered according to Eq. \ref{eq:knight} when $\mathbf{d}_\vk || \mathbf{H}$. This is 
valid as long as the $\mathbf{d}_\vk$ remains 'pinned' in the material due to \eg spin-orbit coupling, and hence does not rotate 
with $\mathbf{H}$. Otherwise, the Knight shift would remain unaltered in any direction. Therefore, an anisotropic Knight shift 
is a strong signature of a vector character of the superconducting order parameter, and hence
of a spin-triplet superconducting state. 
\par
\begin{figure}[h!]
\centering
\resizebox{0.45\textwidth}{!}{
\includegraphics{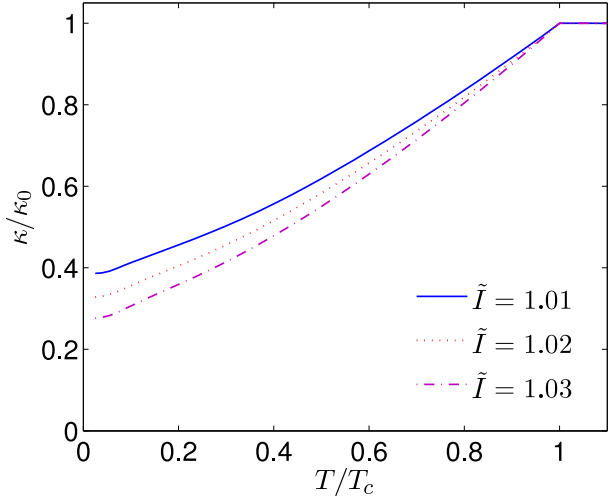}}
\caption{(Color online) Knight shift for a ferromagnetic superconductor, using self-consistently obtained order parameters, for 
three different values of $\tilde{I}$. The field is here applied $\mathbf{H}\parallel \mathbf{d}_\vk$.}
\label{fig:knight}
\end{figure}

In Fig. \ref{fig:knight}, we plot the Knight shift for several values of $\tilde{I}$. It is interesting to note that $\kappa(0)$ is 
reduced with increasing $\tilde{I}$. Physically, this may be understood by realizing that the density of states of ungapped minority 
spins at the Fermi level decreases as the exchange splitting between the majority- and minority bands increases. This results directly in 
a lower amount of polarizable quasiparticles, and hence the Knight shift becomes suppressed. For a fully polarized ferromagnet 
(half-metal), the Knight shift would therefore be identical to an $s$-wave singlet superconductor for an applied field satisfying 
$\mathbf{H}\parallel \mathbf{d}_\vk$. This fact emphasizes the importance of measuring the spin susceptibility along several 
directions to identify the proper spin-symmetry of the superconductor.
\par
As a final experimental probe for the interplay between ferromagnetism and superconductivity, we employ a Blonder-Tinkham-Klapwijk 
formalism \cite{BTK} to calculate the tunneling between a normal metal and a ferromagnetic superconductor in the clean limit, using the 
self-consistently obtained values of the order parameters in the problem. From the results of Ref.~\onlinecite{linder_gronsleth}, 
we find that the normalized tunneling conductance may be written as
\begin{align}
\frac{G}{G_0} &=\sum_\sigma \int^{\pi/2}_{-\pi/2} \text{d}\theta \cos\theta [1 + |r_\sigma^A(eV,\theta)|^2 - |r_\sigma^N(eV,\theta)|^2],
\end{align}
where $G_0$ is the normal-state conductance. Above, $r_\sigma^A(eV,\theta)$ and $r_\sigma^N(eV,\theta)$ designate the Andreev- and normal-reflection coefficient, respectively, and read
\begin{widetext}
\begin{align}\label{eq:cof1}
r_\sigma^N &= -1 + \frac{2k_F\cos\theta[u_\sigma(\theta_{s+}^\sigma)u_\sigma(\theta_{s-}^\sigma)(\Upsilon_+^\sigma)^*+v_\sigma(\theta_{s-}^\sigma)v_\sigma(\theta_{s+}^\sigma)\gamma_\sigma(\theta_{s-}^\sigma)\gamma_\sigma^*(\theta_{s+}^\sigma)(\Upsilon_-^\sigma)^* ]}{u_\sigma(\theta_{s+}^\sigma)u_\sigma(\theta_{s-}^\sigma)|\Upsilon_+^\sigma|^2 - v_\sigma(\theta_{s-}^\sigma)v_\sigma(\theta_{s+}^\sigma) \gamma_\sigma(\theta_{s-}^\sigma)\gamma_\sigma^*(\theta_{s+}^\sigma)|\Upsilon_-^\sigma|^2.}, \notag\\
r_\sigma^A &= \frac{4k_F\cos\theta q^\sigma\cos\theta_s^\sigma v_\sigma(\theta_{s+}^\sigma)u_\sigma(\theta_{s-}^\sigma)\gamma_\sigma^*(\theta_{s+}^\sigma)}{u_\sigma(\theta_{s+}^\sigma)u_\sigma(\theta_{s-}^\sigma)|\Upsilon_+^\sigma|^2 - v_\sigma(\theta_{s-}^\sigma)v_\sigma(\theta_{s+}^\sigma) \gamma_\sigma(\theta_{s-}^\sigma)\gamma_\sigma^*(\theta_{s+}^\sigma)|\Upsilon_-^\sigma|^2.}.
\end{align}
\begin{figure}[h!]
\centering
\resizebox{1.0\textwidth}{!}{
\includegraphics{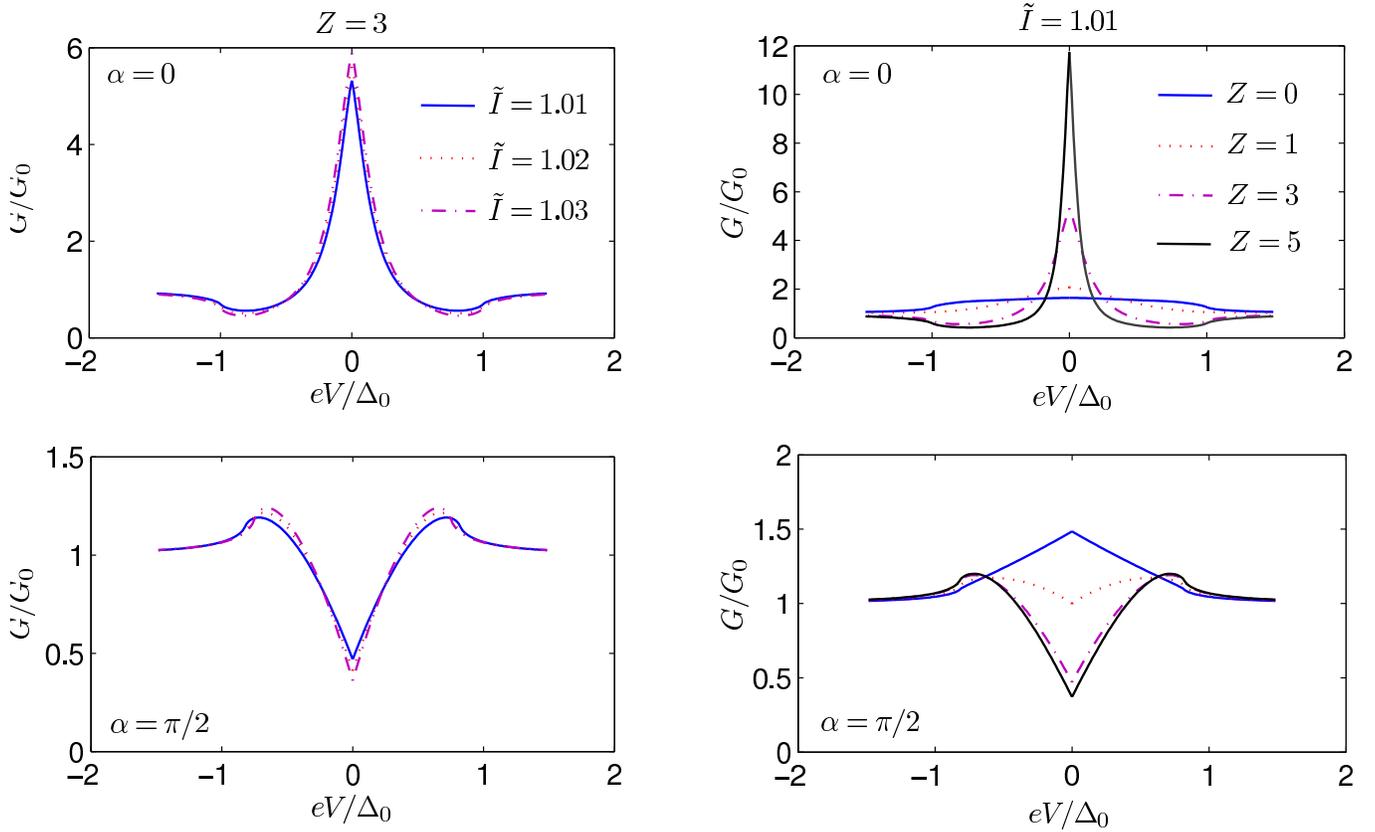}}
\caption{(Color online) Plot of the tunneling conductance of a normal/ferromagnetic superconductor junction for $\alpha=0$ 
and $\alpha=\pi/2$, using self-consistently obtained solutions at $T=0$. In the left column, we fix the tunneling barrier strength $Z=2mV_0/k_F=3$ and plot the conductance for several values of the Stoner interaction $\tilde{I}$. In the right column, we fix $\tilde{I}=1.01$ and plot the conductance for several values of $Z$.}
\label{fig:conductance}
\end{figure}
\end{widetext}
We have defined $Z=2mV_0/k_\mathrm{F}$ as a measure of the barrier strength, where $m$ is the quasiparticle mass, $V_0$ is the scattering 
strength of the barrier, and $k_\mathrm{F}$ is the Fermi momentum. Moreover, $\theta$ is the angle of incidence of incoming electrons 
from the normal side and we have implicitly incorporated conservation of group velocity and conservation of momentum parallel to the 
barrier, \ie $k_\text{F}\sin\theta = q^\sigma\sin\theta_s^\sigma$. Finally, we have introduced
\begin{equation}
\Upsilon_\pm^\sigma=q^\sigma\cos\theta_{s}^\sigma\pm k_F\cos\theta \pm \i k_FZ
\end{equation}
and $\gamma_\sigma(\theta) = \Delta_\sigma(\theta)/|\Delta_\sigma(\theta)|$, 
$\theta_{s+}^\sigma = \theta_s^\sigma$, 
$\theta_{s-}^\sigma = \pi-\theta_s^\sigma.$ In the quasiclassical approximation $E_F \gg (\Delta_0,\varepsilon)$, the wave-vectors read 
\begin{align}
k_\text{F} = \sqrt{2mE_F},\; q^\sigma = \sqrt{2m(E_F + \sigma IM)}
\end{align}
while the spin-generalized coherence factors are
\begin{align}
u_\sigma(\theta_{s\pm}^\sigma) &=\frac{1}{\sqrt{2}}\{1+\sqrt{1 - (|\Delta_\sigma(\theta_{s\pm}^\sigma)|/E)^2}\}^{1/2},\notag\\
v_\sigma(\theta_{s\pm}^\sigma) &= \frac{1}{\sqrt{2}}\{1-\sqrt{1 - (|\Delta_\sigma(\theta_{s\pm}^\sigma)|/E)^2}\}^{1/2}.
\end{align}
In Fig. \ref{fig:conductance}, we plot the conductance spectra of a normal/ferromagnetic superconductor junction. By writing the gap 
as $\Delta = \Delta_0\cos(\theta-\alpha)$, we allow for an arbitrary orientation of the gap with respect to the crystallographic axes. 
The features seen in the conductance spectra are qualitatively different for $\alpha=0$ and $\alpha=\pi/2$. In the first case, the 
electron- and hole-like quasiparticles entering the superconductor experience a constructive phase-interference which gives rise to 
the formation of a zero-energy state that is bound to the surface of the superconductor. The resonance condition for the formation 
of such zero-energy states is $\Delta(\theta) = -\Delta(\pi-\theta)$, \cite{hu} and the bound states are manifested as a giant peak 
in the zero-bias conductance \cite{tanaka}. Note that such states exist even if the spatial depletion of the superconducting order 
parameter is not taken into account, which may be shown analytically \cite{lofwander}. Taking into account the reduction the gap 
experiences close to the interface compared to its bulk value, is known to yield the same qualitative features as the usual 
step-function approximation, with the exception of additional, smaller peaks at finite bias voltages due to non-zero bound 
states \cite{barash}. From Fig. \ref{fig:conductance}, we see that the effect of increasing the exchange field amounts to 
sharper features in the conductance spectra. With increasing $\tilde{I}$, the zero-bias conductance peak becomes larger for 
$\alpha=0$, while the dip structure for $\alpha=\pi/2$ becomes more pronounced. Physically, this may be understood by the 
increased contribution from majority spin carriers. The contribution from the minority spin carriers is constant for the 
entire low-energy regime, and leads to less pronounced features in the conductance. The effect of the barrier strength $Z$ is seen in the left column of Fig. \ref{fig:conductance}. For $\alpha=0$, increasing $Z$ leads to a higher peak at zero bias, while increasing $Z$ suppresses the conductance for $\alpha=\pi/2$. 
\par
It is also worth emphasizing the relation between the tunneling conductance and the bulk DOS of the superconductor. As is well-known, the conductance of a normal/$s$-wave superconductor junction in the tunneling limit approaches the DOS of the bulk superconductor \cite{BTK}. The same argument is valid for a $d_{x^2-y^2}$-wave superconductor \cite{tanaka}. One might be tempted to conclude that the tunneling conductance will always approach the bulk DOS of the superconductor in the strong barrier limit as long as there is no formation of zero-energy states. However, closer examination reveals that this is not necessarily so. 
\par
To illustrate this, we draw upon some results obtained in Ref.~\onlinecite{kashiwayaPRB}. In general, the conductance of an N/S junction in the tunneling limit may be written as
\begin{align}\label{eq:g}
G(eV) \approx \frac{\int^{\pi/2}_{-\pi/2} \text{d}\theta_N\sigma_N\cos\theta_N\rho_S(eV)}{\int^{\pi/2}_{-\pi/2} \text{d}\theta_N\sigma_N\cos\theta_N},
\end{align}
where $\sigma_N$ is the normal-state conductance for a given angle of incidence $\theta_N$ and $\rho_S$ is the surface DOS for the superconductor. In the absence of zero-energy states, the surface DOS coincides with the bulk DOS of the superconductor, i.e. $\rho_S = \rho_0$, where
\begin{align}
\rho_0(eV) = \int^{\pi/2}_{-\pi/2} \text{d}\theta_N \text{Re}\Big\{ \frac{eV}{\sqrt{eV^2-|\Delta(\theta_N)|^2}}\Big\}.
\end{align}
An important consequence of the above equation is that the tunneling conductance may be interpreted as the expectation value of $\rho_S$ with a weighting factor $\sigma_N\cos\theta_N$. 
\par
\begin{figure}[h!]
\centering
\resizebox{0.45\textwidth}{!}{
\includegraphics{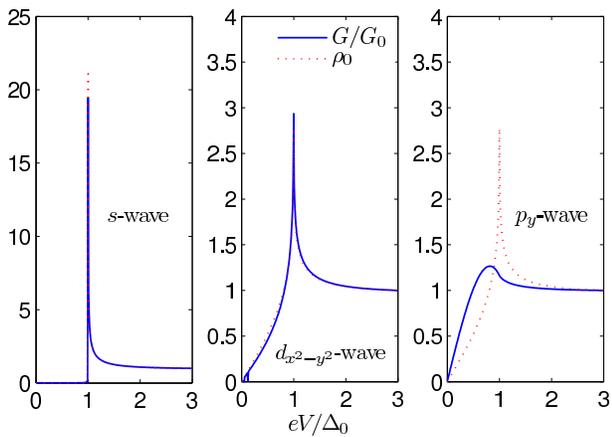}}
\caption{(Color online) Plot of the normalized conductance $G/G_0$ and bulk DOS $\rho_0$ for three different symmetries of the superconducting state in the tunneling limit. Only in the $p_y$-wave case is there a difference between these two quantities.}
\label{fig:compare}
\end{figure}
\begin{figure}[h!]
\centering
\resizebox{0.45\textwidth}{!}{
\includegraphics{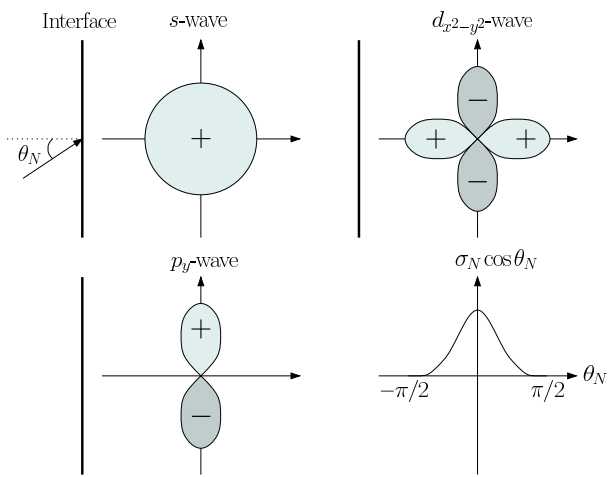}}
\caption{(Color online) Illustration of the different symmetry states considered here and a qualitative sketch of the weighting factor $\sigma_N\cos\theta_N$.}
\label{fig:explanation}
\end{figure}

Let us now compare three different superconducting symmetries to illustrate the relation between the conductance and the DOS. We consider an $s$-wave, $d_{x^2-y^2}$-wave, and $p_y$-wave symmetry, none of which feature zero-energy surface states (Fig. \ref{fig:explanation}). Naively, one might therefore expect that the conductance should converge towards $\rho_0$ in the tunneling limit. However, it turns out that the weighting factor $\sigma_N\cos\theta_N$, which is peaked around $\theta_N=0$, plays a major role in this scenario. In Fig. \ref{fig:compare}, we plot both the tunneling conductance $G(eV)/G_0$ and the bulk DOS $\rho_0$ for these three symmetries and fix $Z=20$. We regain the well-known results that $G(eV)/G_0 \to \rho_0$ for large $Z$ in the $s$-wave and $d_{x^2-y^2}$-case. However, the conductance and DOS differ in the $p_y$-wave case.
\par
The reason for the deviation between $G/G_0$ and $\rho_0$ in the $p_y$-wave case may be understood by consulting Fig. \ref{fig:explanation}. As seen, the weighting factor is peaked around normal incidence $\theta_N=0$. In the $s$-wave and $d_{x^2-y^2}$-case, the gap magnitude is maximal at $\theta_N=0$ and replacing the weighting factor in Eq. (\ref{eq:g}) with unity has little or no consequence. The situation is dramatically different in the $p_y$-wave case. Now, the gap magnitude is actually \textit{zero} for normal incidence, and it is precisely this contribution that will dominate the integration over angles in Eq. (\ref{eq:g}). Therefore, replacing the weighting factor with unity, in order to obtain the DOS, has a non-trivial consequence in the $p_y$-wave case. This analysis illustrates how the conductance and bulk DOS in the absence of zero-energy states are not always the same in the tunneling limit. Note that the orientation of the interface with respect to the symmetry of the order parameter is crucial with regard to the measured conductance spectra and the surface DOS. For instance, even at $\alpha=\pi/4$ there is an appearance of a large zero-bias conductance peak for the $p$-wave pairing considered here, although the gap orientation does not satisfy the condition for perfect formation of zero-energy states.

\section{Discussion}\label{sec:discuss}
We have discussed a mean-field model where itinerant ferromagnetism coexists with non-unitary, triplet superconductivity, 
with a gap that contains line nodes. The precise symmetry of the order parameter in the ferromagnetic superconductors UGe$_2$, URhGe, 
UCoGe is still under debate, although most experimental findings and theoretical considerations strongly point towards the realization 
of a triplet superconducting order parameter. It is plausible that such a superconducting order parameter is 
non-unitary, thus breaking time-reversal symmetry in the spin channel of the Cooper pair. 
\par
The orbital symmetry of the superconducting order parameter in ferromagnetic superconductors is a more subtle issue. In 
Ref. \onlinecite{linder}, a mean-field model for isotropic, chiral $p$-wave gaps in a background of itinerant ferromagnetism 
was constructed. In that work, pairing was assumed to occur both for majority- and minority-spins, resulting in for instance 
a double-jump structure in the specific heat capacity. An isotropic, chiral $p$-wave order parameter has a constant magnitude, 
which is favorable in terms of maximizing the condensation energy gained in the superconducting state. Assuming an isotropic 
density of states at the Fermi level and a separable pairing potential of the form $V_{\vk\vk'} = -g\lambda_\vk\lambda_{\vk'}$, 
the condensation energy gained at $T=0$ in the superconducting state reads
\begin{equation}
E = -\frac{N(0)\Delta_0^2}{2}\langle |\lambda_\vk|^2 \rangle, 
\end{equation}
where $\Delta_0$ is the maximum value of the gap and $\langle \ldots \rangle$ denotes the angular average over the Fermi surface. 
This clearly shows the advantage of an isotropic gap $|\lambda_\vk|=1$. The general principle is well-known: the system prefers 
to have the Fermi surface as gapped as possible. However, factors such as spin-orbit pinning energy and lattice structure may 
conspire to prevent a fully isotropic gap. We also note that in our model, the ferromagnetic ordering enters at a much higher 
temperature than the superconducting order unless $\tilde{I}$ is very close to unity. This is consistent \cite{saxena, aoki} 
with the experimental findings for the ratio between the critical temperatures for ferromagnetic and superconducting order,
 $T_c^\text{FM}/T_c^\text{SC}$, except for UCoGe where the ratio is $\simeq 3$.\cite{huy}
\par
The experiments performed so far are indicative of a single gap, or at
least a strongly suppressed second gap, in the ferromagnetic
superconductors. For instance, no double-jump features have been
observed in the specific heat capacity \cite{saxena} for UGe$_2$. This
warrants the investigation of a single-gap model, possibly with line
nodes as suggested by Harada \etal.\cite{harada} 
Theoretically, the absence of the SC gap in the minority spin subband
can be justified by considering the effect of Zeeman splitting on the
electronic density of states (see discussion in Sec.~\ref{sec:theory}
and Ref.~\onlinecite{nevidomskyy}). In general, it should be possible to 
discern the presence of two gaps by analyzing specific heat or point-contact 
spectroscopy measurements, unless one of the gaps is very small. 
\par
Apart from this, another possible scenario, specific to UGe$_2$, can
be invoked to explain the observed gapless behaviour in the minority
spin subband. This is the \emph{meta-magnetic transition} that occurs inside
the FM phase of UGe$_2$ and separates the two ferromagnetic phases
with different values~\cite{tateiwa, uhlarz} of magnetization $M$.
The reason this meta-magnetic transition in UGe$_2$ is of great
importance is because the specific heat measurements clearly
indicate~\cite{tateiwa} that the maximum of superconducting $T_c$
occurs not at the FM to PM transition, but at some lower pressure
$p_x\approx 12$~kbar that coincides precisely with the meta-magnetic
transition~\cite{harada,uhlarz}. 

One can think of this transition as a point where
the value of low-temperature magnetization $M$ sustains a jump. While
the  microscopic origin of this transition is not known, an idea has
been put forward~\cite{sandeman} that it may be due to a sharp change in the density
of states (DOS) due to the existence of a double peak in its structure close
to the Fermi level. What happens according to this
scenario is that applied pressure makes the Fermi level ``sweep
through'' the double-peak structure in the DOS, thereby sharply
increasing the density of states in the majority spin channel. It
follows from a simple Stoner instability argument that such an increase
in the DOS would lead to a larger value of effective interaction
$\tilde{I}\equiv I N(0)$ and thus higher magnetization $M$. But this
also means that the ratio of the DOS in the two spin channels,
$N_\uparrow/N_\downarrow$, sharply increases at the meta-magnetic transition.
It follows from Eqs.~(\ref{eq:gap}, \ref{eq:gap1}, \ref{eq:Tc}) that the ratio between the SC gaps in
the two spin subbands
\be
\frac{\Delta_\downarrow}{\Delta_\uparrow}\propto \frac{T_c^\downarrow}{T_c^\uparrow}
= \frac{\exp(-1/gN_\downarrow)}{\exp(-1/gN_\uparrow)}
\ee
thus becomes very small, justifying the assumption
$\Delta_\downarrow=0$ made in this work.

We note in passing that from an experimental point of view, a
complication with UGe$_2$ is that the superconductivity does not
appear at ambient pressure, in contrast to URhGe and UCoGe. The
necessity of considerable pressure restricts the use of certain
experimental techniques, and this is clearly a challenge in terms of
measuring for instance conductance spectra of UGe$_2$. Another
experimental quantity which would be of high interest to obtain from
for instance \textit{ab initio} calculations, is the thermal expansion
coefficient, which may be directly probed in high-pressure experiments \cite{sato}.

We also underline that in our model the magnetism is assumed to coexist uniformly with superconductivity. 
Depending on the geometry of the sample, it is likely that the intrinsic magnetization gives rise to a 
self-induced vortex phase. In a thin-film structure where the thickness $t$ is smaller than the vortex 
radius $\lambda$, we expect that ferromagnetism and superconductivity may be realized in a vortex-free 
phase, similarly to a thin-film $s$-wave superconductor in the presence of an in-plane magnetic field. 
Further refinements leading to a more realistic model of a ferromagnetic superconductor should include 
the presence of spin-orbit coupling, which inevitably is present in heavy-fermion superconductors, in 
addition to the presence of vortices. Nevertheless, we believe that our model should capture important 
qualitative features of how the interplay between ferromagnetism and superconductivity may be manifested 
in experimentally accessible quantities. In particular, experiments on transport properties of ferromagnetic 
superconductors, such as the Josephson current and point-contact spectroscopy, would be of high interest 
to further illucidate the pairing symmetry realized in ferromagnetic superconductors. 

\section{Summary}\label{sec:summary}
In conclusion, we have constructed a mean-field theory of triplet superconductivity in the background of itinerant 
ferromagnetism, where the superconducting order parameter contains line nodes and the minority spin band remains 
ungapped at the Fermi level. We have solved the self-consistency equations for the order parameters in the problem, and find that 
ferromagnetism enhances superconductivity, while the ferromagnetism itself is virtually unaffected by the 
presence of superconductivity. We have made several predictions for experimentally accessible quantities: heat 
capacity, Knight shift, and tunneling conductance spectra. Our results may be helpful in the interpretation of experimental 
data, and could provide tools concerning the issue of identifying the pairing symmetry of ferromagnetic superconductors.
\acknowledgments
J.L. wishes to express his gratitude to Y. Tanaka at Nagoya University for his hospitality, where parts of this 
work were completed. N. Sato is also thanked for very useful discussions.
This work was
  supported by the Norwegian Research Council Grants No. 158518/431
  and No. 158547/431 (NANOMAT), and Grant No. 167498/V30 (STORFORSK).


\begin{thebibliography}{99}
\bibitem{footnote-ZrZn2} Ironically, Fay and Appel\cite{fay-appel} have developed
their theory with ZrZn$_2$ as a potential application in mind, but it
took further 20 years before superconductivity was finally discovered in this
compound~\cite{pfleiderer}.
\bibitem{footnote-URhGe} Although the superconductivity is observed at ambient pressure in 
URhGe\cite{aoki}, UCoGe\cite{huy} and ZrZn$_2$\cite{pfleiderer}, it is believed that it occurs 
on the border of ferromagnetism and that one could in principle suppress the Curie temperature 
to zero by applying pressure (sometimes a negative pressure would be required), making these 
compounds similar in this respect to UGe$_2$, where SC is observed~\cite{saxena, pfleiderer} 
in the vicinity of the pressure-tuned phase transition.
\bibitem{footnote-orbital} Notice that time-reversal symmetry may be spontaneously broken in 
the orbital part (angular momentum) of the Cooper pair wavefunction even if the state is unitary.
\end{thebibliography}
\end{document}